\documentstyle[twocolumn,aps,prl]{revtex}
\input epsf
\begin{document}
\draft

\title{Critical Collapse of Skyrmions}

\author{Piotr Bizo\'n${*}$}
\address{Department of Mathematics, University of Michigan, Ann Arbor,
 Michigan}
\author{Tadeusz Chmaj}
\address{Institute of Nuclear Physics, Cracow, Poland}

\date{\today}

\maketitle

\begin{abstract}
We study first order phase transitions in the gravitational collapse of spherically
symmetric skyrmions. 
Static sphaleron solutions are shown to play the role of critical
solutions separating black-hole spacetimes from no-black-hole
spacetimes. In particular, we find a new type of first order phase transition
where subcritical data do not disperse but evolve towards a static regular
 stable solution.
 We also demonstrate explicitly that the  near-critical solutions 
depart from
the intermediate asymptotic regime along the  unstable manifold of the
critical solution.

\end{abstract}

\narrowtext

The understanding of the dynamics of gravitational collapse is a major
theoretical challange in classical general relativity. A few years ago a new
twist into this problem was provided by Choptuik's numerical discovery
of critical behaviour in the gravitational collapse of spherically symmetric
massless scalar field~\cite{matt}. Analysing the evolution of one-parameter
families of regular initial data, Choptuik showed that the transition between
collapse and dispersion has a universal character which is reminiscent of
a second order phase transition with the continuous change of black-hole mass.
Subsequently, similar critical behaviour has been observed in several other
models of gravitational collapse, and significant progress has been made towards
the analytical understanding of the numerical phenomenology
(see~\cite{carsten} for a recent review and bibliography).

Recently, a qualitatively new type of critical collapse was found in the
Einstein-Yang-Mills system~\cite{eym}, where, for some initial data, the
fundamental Bartnik-McKinnon sphaleron (static soliton with one unstable mode)
plays the role of a critical solution. In this case there is a gap in the
spectrum of black-hole masses which is reminiscent of the first order
phase transition.

Surprisingly, none of the models studied so far had a {\em stable} static
solution (a "star") and consequently the possible endpoints of  evolution
were basically reduced to the collapse/dispersion 
alternative. As far as we know, the only exception is the collapse/oscillaton
alternative recently found in the collapse of massive scalar
field~\cite{brady}.
If a model has a star-like solution, then there will be an open set of initial
data which evolve towards this solution (basin of attraction) and it is 
natural to expect the
occurence of critical behaviour at the boundary of this set.
 
The Einstein-Skyrme model which possesses the whole
zoo of static regular solutions, both stable and unstable, seems to be a
suitable testing ground to address this issue.
In this letter we focus our attention on a new type of first order phase
transition we found in this model for baryon-number-one data. We also briefly
discuss a novel feature of mass gap in the baryon-number-zero sector.
A detailed description of both first and
second order phase transitions will be published
elsewhere~\cite{bct}.

We consider the Einstein-Skyrme system, so the matter in our model is an
$SU(2)$-valued scalar function $U(x)$ (called a chiral field) with dynamics
given by the Lagrangian
\begin{eqnarray}
\!L&=& \frac{f^2}{4}\,Tr(\nabla_a \nabla^aU^{-1})\!
+\!\frac{1}{32 e^2}\, Tr[(\nabla_aU)U^{-1},(\nabla_bU)U^{-1}]^2,
\end  {eqnarray}
where $\nabla_a$ is the covariant derivative with respect to the spacetime
metric. The two coupling constants $f$ and $e$ are hereafter set to one,
which amounts to using $f/e$ and $1/ef$ as the units of mass and length,
respectively.

We specialize to spherical symmetry. For the metric we use the 
polar time slicing and the areal radial coordinate
\begin{eqnarray}
ds^2 &=& -e^{-2\delta (r,t)} N (r,t) dt^2 + N^{-1} (r,t) dr^2 + r^2 d \Omega^2.
\label{METRIC}
\end{eqnarray}
For the chiral field we assume the hedgehog ansatz $U=exp(i\vec\sigma \cdot
\hat r F(r,t))$, where $\vec\sigma$ is the vector of Pauli matrices.
Using overdots and primes to denote $\partial / \partial t$ and 
$\partial / \partial r$ respectively, we introduce an auxilary variable
$P=u e^{\delta} N^{-1} {\dot F}$, where $u=r^2+2 \sin^2{F}$.
Then, the Einstein-Skyrme equations reduce to
\begin{eqnarray}
\!{\dot F} &=& e^{-\delta} N \frac{P}{u},
\label{fDOT}
\\
\!{\dot P} &=& (e^{-\delta} N u F')' \!
+ \!\sin(2F) e^{-\delta} \left( N ( \frac{P^2}{u^2} \!- \!F'^2) \!
- \!\frac{u}{2 r^2}\! - \!\frac{1}{2}\right),
\\
\!{\dot N} &=& -\frac{2 \alpha}{r} e^{-\delta} N^2 P F',
\label{LAPSE}
\\
\!{\delta'} &=& -\frac{\alpha u}{r} \left( \frac{P^2}{u^2} + F'^2 \right),
\\
\!N' &=& \frac{1-N}{r} - \frac{\alpha}{r} 
\left( 2 \sin^2{\!F}+\frac{\sin^4{\!F}}{r^2}+
 u N (\frac{P^2}{u^2} + F'^2)\right).
\label{HAM}
\end{eqnarray}
Here $\alpha=4\pi G f^2$ is the dimensionless coupling constant. In order to
solve
the initial value problem numerically we have implemented a free evolution
scheme using a second order finite difference method on uniform grid. The
conservation of the hamiltonian constraint (7) was used only to check the accuracy
of the code. To ensure regularity at the center we impose the boundary
condition  $F(r,t)=O(r)$ for $r \rightarrow 0$. 
To minimize reflections from the outer boundary of the grid we impose a sort of
outgoing wave condition there.
Asymptotic flatness requires that  $F(r,0)=B \pi + O(1/r^2)$ at infinity,
where the integer $B$, called the baryon number, is equal to the topological
degree of the chiral field. As long as no horizon forms, the baryon number
of initial data is automatically preserved during the evolution.

Below we describe our results for regular initial data with baryon number
one and zero.

$B=1:$ Let us recall~\cite
{my} that in this case, for $\alpha < \alpha_1
 \simeq 0.040378$, the 
Eqs.(3-7) have two regular static solutions (solitons) which we denote below
by $X^s$ and $X^u$ with $X$ standing for $(F,P,N,\delta$).
The solution $X^s$ is linearly stable - it is the skyrmion distorted by
gravity. The second solution $X^u$ has one linearly unstable mode.
These two solutions coalesce at the bifurcation point $\alpha=\alpha_1$,
and disappear for $\alpha>\alpha_1$. For given $\alpha$
the mass $m_u$ of the unstable soliton
is larger than the  mass $m_s$ of the stable soliton
(for example, for $\alpha=0.02$ we have $m_s \simeq
68.05 $ and $m_u \simeq 80.70 $). As $\alpha \rightarrow 0$,
$m_s$ tends (from below) to the mass of the flat-space skyrmion 
($\simeq 72.92 $), whereas $m_u \rightarrow \infty$, which is the
 manifestation of a nonperturbative character of the solution $X^u$.

Since for $B=1$ the dispersion of the chiral field to infinity is forbidden,
generic initial data end up either as black holes or as the stable soliton $X^s$.
The respective basins of attraction of these two final states depend on the
coupling constant $\alpha$. This is illustrated in Fig.~1 for a typical
time-symmetric kink-type initial data.
%%%%%%%%%%%%%%%%%%%%%%%%%%%%%%%%%%%%%%%%%%%%%%%%%%%%%%%%%%%%%%%%%%%%%%%%
\begin{figure}
\epsfxsize=8cm
\centerline{\epsffile{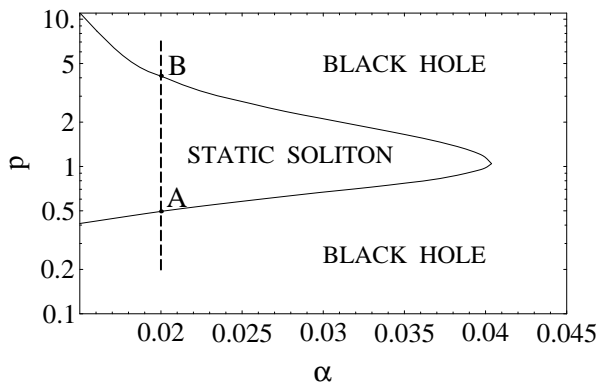}}
\caption{Basins of attraction of the one-parameter family of $B=1$ 
initial data $F_p(r,0)=\pi \tanh(r/p)$
 as a function of the coupling constant $\alpha$. Critical initial data are
 represented by the solid line curve.
 Initial data lying in the
  region enclosed by the critical curve and the $p$-axis evolve towards the 
  skyrmion $X^s$, while initial data
  lying outside that region form a black hole. 
As $\alpha \rightarrow 0$ (not shown in the figure) the basin of attraction 
of $X^s$ tends to the whole positive $p$-axis. 
}

\label{FIG1}
\end{figure}
%%%%%%%%%%%%%%%%%%%%%%%%%%%%%%%%%%%%%%%%%%%%%%%%%%%%%%%%%%%%%%%%%%%%%%%%
 Notice that for $\alpha < \alpha_1$ there are two critical configurations
along the one-parameter family $F_p$
(the points $A$ and $B$ on the dashed line segment on Fig.~1). Since the 
parameter $p$ (the width of a kink)
is inversely proportional to the "condensation" of initial mass,
the transition at the point $A$ could be interpreted as a "weak/strong"
field transition. However at the point $B$ it is the "weaker" configuration
that collapses, simply because it is unable to get rid of the excess mass by the time it shrinks to size
of the stable soliton. 
This behaviour indicates that it is difficult to formulate
a criterion for collapse in terms of initial data alone.

The critical
initial data lying on the boundary  of the basin of attraction of $X^s$
asymptote to the unstable solution $X^u$. In other words, the codimension one
stable manifold of the solution $X^u$ divides (locally) the phase space
into collapsing and non-collapsing data.
 Near-critical
data approach $X^u$, stay in its vicinity for some time, and eventually
collapse to a black hole or decay into $X^s$. This is shown in Fig.~2 for 
a marginally subcritical data (lying 
slightly below the point $B$ on Fig.~1).
%%%%%%%%%%%%%%%%%%%%%%%%%%%%%%%%%%%%%%%%%%%%%%%%%%%%%%%%%%%%%%%%%%%%%%%%
\begin{figure}
\epsfxsize=8cm
\centerline{\epsffile{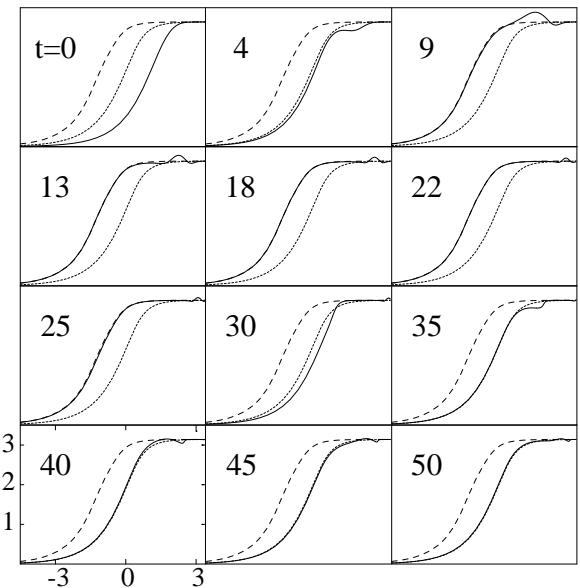}}
\caption{A sequence of profiles of $F(r,t)$ (solid lines) versus $\ln(r)$
from a marginally subcritical evolution of the $B=1$ initial data.
The stable and unstable static solutions are represented by the dotted and
dashed lines, respectively. Here $t$ is the proper time at spatial infinity
(we use the normalization $\delta(\infty,t)=0$).
}
\label{FIG2}
\end{figure}
%%%%%%%%%%%%%%%%%%%%%%%%%%%%%%%%%%%%%%%%%%%%%%%%%%%%%%%%%%%%%%%%%%%%%%%%
During the final stage of the evolution of a marginally supercritical solution
(i.e. when the solution runs away from the vicinity of the critical solution
and forms a black hole) 
almost no energy is
radiated away to infinity.
Therefore the mass gap in this first order phase transition equals $m_u$
(up to $0.1\%$).

For subcritical data the process of settling down to the stable soliton
seems to have a universal late-time behaviour which is dominated by
a fundamental quasinormal mode. The parameters of this mode depend strongly
on $\alpha$. Having convinced ourselves that these parameters are indeed
universal, that is do not depend on the excitation, we have determined them
from the evolution of finite perturbations of $X^s$ (see Fig.~3).
 The details of such perturbation are
washed out very rapidly and a characteristic exponentially damped "ringing"
dominates the late time evolution. 
%%%%%%%%%%%%%%%%%%%%%%%%%%%%%%%%%%%%%%%%%%%%%%%%%%%%%%%%%%%%%%%%%%%%%%%%
\begin{figure}
\epsfxsize=8cm
\centerline{\epsffile{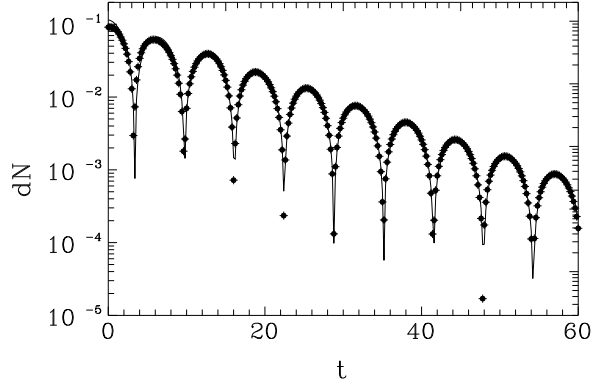}}
\caption{Quasinormal ringing about the stable soliton for $\alpha=0.03$.
We plot $dN=|N(r_0,t)-N^s(r_0)|$ from the evolution of
initial data $F(r,0)=\pi \tanh(r)$ (dots). Here $r_0=0.8$.
The least squares fit for the times $t>10$ (represented by the solid line)
gives the frequency $\omega \simeq 0.494$ and the damping rate $\tau \simeq
11.69$.
}
\label{FIG3}
\end{figure}
%%%%%%%%%%%%%%%%%%%%%%%%%%%%%%%%%%%%%%%%%%%%%%%%%%%%%%%%%%%%%%%%%%%%%%%%

The above results are in perfect agreement with linear~\cite{hds,my} and
nonlinear~\cite{zhou} stability analysis of gravitating skyrmions.
In the intermediate asymptotics the evolution of near-critical data is well
approximated by the linearization about $X^u$
\begin{equation}
\!X_p(r,t) \approx  X^u(r) \!+ \! C (p-p^{\star}) e^{\lambda t} 
\delta X^u(r)
+ {\mathrm{decaying}\atop \mathrm{modes}},
\end{equation}
where $\delta X^u(r)$ is the single unstable eigenmode associated with a
positive 
eigenvalue $\lambda$. 
Depending on whether $p>p^{\star}$ or $p<p^{\star}$, the solution $X_p$ 
eventually collapses to a black hole or decays to $X^s$. The "lifetime" $T$ of
 a near-critical
solution staying in the vicinity of $X^u$ is determined by the 
time in which the amplitude of the unstable mode grows to a finite size:
$|p-p^{\star}| e^{\lambda T} \sim O(1)$, which gives $T \sim 
-\lambda^{-1} \ln|p-p^{\star}|$.
Thus, the larger $\lambda$, the better fine-tuning is required to see the
solution $X^u$ clearly pronounced as the intermediate attractor.

We have verified the formula (8) by comparing the unstable eigenmode 
$\delta X^u$ with the snapshots of a near-critical solution departing from the 
vicinity of $X^u$. This is shown in Fig.~4 for the same initial data as in
Fig.~2. For $t=24$ the profile of
$\dot F(r,24)$ is practically indistinguishable from the (suitably normalized)
unstable eigenmode $\delta F^u(r)$. A slight deviation of $\dot F(r,23)$
from $\delta F^u(r)$ for large $r$ is due to the fact that by $t=23$ the 
decaying  modes 
in (8) have not yet died away
completely. On the other hand, for $t=25$ the deviation for small $r$ signals 
the onset of nonlinear regime.
%%%%%%%%%%%%%%%%%%%%%%%%%%%%%%%%%%%%%%%%%%%%%%%%%%%%%%%%%%%%%%%%%%%%%%%%
\begin{figure}
\epsfxsize=8cm
\centerline{\epsffile{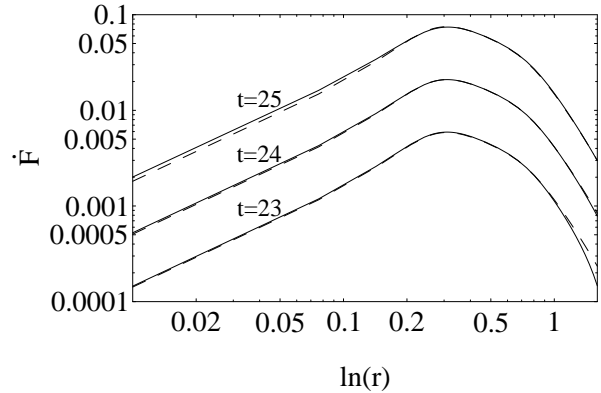}}
\caption{
The snapshots of $\dot F(r,t)$ (solid lines) departing from the vicinity of the
critical solution $X^u$. On each snapshot we superimpose (dashed lines) the 
unstable eigenmode
$\delta F^u(r)$ computed via linear perturbation analysis. For $t=24$
the amplitude of
$\delta F^u(r)$ is normalized to the maximum of $\dot F(r,24)$. The
corresponding amplitudes for $t=23$
and $t=25$ are rescaled by the factors $e^{-\lambda}$ and $e^{\lambda}$,
respectively.
}
\label{FIG4}
\end{figure}
%%%%%%%%%%%%%%%%%%%%%%%%%%%%%%%%%%%%%%%%%%%%%%%%%%%%%%%%%%%%%%%%%%%%%%%%%%%%
Using (8), we have also computed the eigenvalue 
$\lambda$ directly from the
nonlinear evolution by monitoring $F_p(r,24)$ at several discrete radii $r_i$,
$i=1,...,n$,
and evaluating the averaged quantity 
$\frac{1}{n\Delta t}\sum \ln \left(\dot F_p(r_i,24+\Delta t)/\dot
F_p(r_i,24)\right)$ for some small $\Delta t$. The result agrees up to three
decimal places with the eigenvalue obtained via linear stability analysis.

$B=0$: 
In this topological sector, the Eqs.(3-7) have a pair of static regular solutions
for $\alpha<\alpha_0 \simeq  0.00147$~\cite{my}. One of them 
has two unstable modes, hence it plays no role in the evolution of generic
initial data.
The second solution, call it $Y^u$, has  one 
unstable mode, so it is a candidate for the critical solution. Indeed,
analysing the ingoing "generalized Gaussian" profile 
$F(r,0)= A r^3 e^{ -(r-r_0)^4/\Delta^4}$, we
have found $Y^u$ as the intermediate attractor at the border between
collapse and dispersion. Since in this case the  critical behaviour is 
analogous
to the first order phase transition in the EYM model~\cite{eym}, here we discuss
 only the issue of mass gap in this transition. Analysing near-supercritical
 solutions we found that, in contrast to the previously studied cases,
 the mass gap is {\em not} equal to the mass of the sphaleron. 
 In order to determine the 
 mass gap more
  precisely we adopted the following strategy.
Rather than improving
the accuracy of fine-tuning (which is computationally time-consuming 
because the eigenvalue of the unstable eigenmode is rather large), we began the evolution
with specially prepared initial data of the form of the static solution
$Y^u$ plus the unstable eigenmode with a small positive amplitude:
$Y(r,0)=Y^u(r)+\epsilon \delta Y^u(r)$.
In other words, we pushed the static solution  
along its unstable manifold and let it collapse. 
It turned out that 
the mass gap is considerably
smaller than the mass of the critical solution (which of course sets the upper
bound for a mass gap). This is illustrated in Fig.~5,
where one can clearly see how a substantial amount of the initial mass is 
being radiated away to infinity.
The resulting black hole has the mass $\simeq  253.1$, which
is $75.6\%$ of the mass of $Y^u$.
%%%%%%%%%%%%%%%%%%%%%%%%%%%%%%%%%%%%%%%%%%%%%%%%%%%%%%%%%%%%%%%%%%%%%%%%
\begin{figure}
\epsfxsize=8cm
\centerline{\epsffile{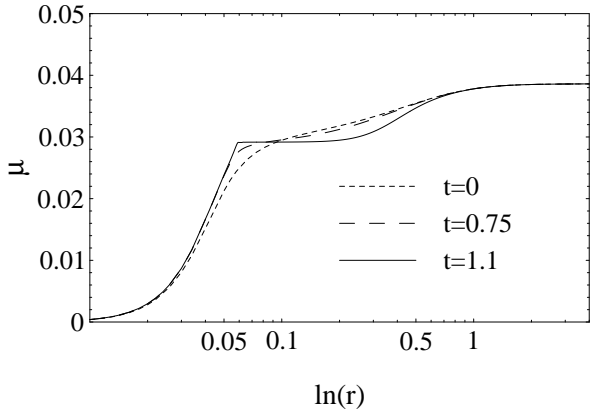}}
\caption{
The mass aspect $\mu(r,t)$ (defined by $N=1-2\mu/r$)
from the evolution of the static solution $Y^u$ perturbed along the unstable
 direction (for $\alpha=0.00145$). The mass inside radius $r$ at
 time $t$ measured in units of $f/e$ is given by
 $m(r,t)=\frac{4\pi}{\alpha} \mu(r,t)$. The plateau of $\mu(r,t)$ determines
 the mass gap.
}
\label{FIG5}
\end{figure}
%%%%%%%%%%%%%%%%%%%%%%%%%%%%%%%%%%%%%%%%%%%%%%%%%%%%%%%%%%%%%%%%%%%%%%%%%%

Concluding, we believe that the results presented here have clarified certain
aspects of first order phase transitions in gravitational collapse and,
 in combination with the results of~\cite{eym} and~\cite{brady}, 
they help to understand which features of this phenomenon are
generic. 

{\em Acknowledgments.} 
This research was supported in part by the KBN grant
PB750/P3/94/06. 
%%%%%%%%%%%%%%%%%%%%%%%%%%%%%%%%%%%%%%%%%%%%%%%%%%%%%%%%%%%%%%%%%%%%%%%%


\begin{references}
\bibitem[*]{}
 On leave of absence from
Institute of Physics, Jagiellonian University, Cracow, Poland.

\bibitem{matt} M. W. Choptuik,
 Phys. Rev. Lett. {\bf 70}, 9-12 (1993).

\bibitem {carsten} C. Gundlach, 
 preprint gr-qc/9712084.

\bibitem{eym} M. W. Choptuik, T. Chmaj, and P. Bizo\'n,
 Phys. Rev. Lett. {\bf77}, 424-427 (1996).

\bibitem{brady} P. R. Brady, C. M. Chambers, and S. M. C. V. Gon\c{c}alves,
 Phys. Rev. {\bf D56}, 6057-6061 (1997).
 
\bibitem{bct} P. Bizo\'n, T. Chmaj, and Z. Tabor, in preparation.
 
\bibitem{my} P. Bizo\'n and T. Chmaj, 
Phys. Lett. {\bf B 297}, 55-61 (1992).

\bibitem{hds} M. Heusler, S. Droz, and N. Straumann,
 Phys. Lett. {B \bf 271}, 61-67 (1991).

\bibitem{zhou} M. Heusler, N. Straumann, and  Z.~Zhou, 
Helv. Phys. Acta {\bf 66}, 615-631 (1993).

\end{references}
\end{document}